\documentclass[7pt]{article}
\twocolumn
\usepackage[utf8]{inputenc}
\usepackage{graphicx}
\usepackage{subcaption}
\usepackage{caption}
\usepackage{siunitx}
\usepackage{hyperref}
\graphicspath{{./images/}}
\usepackage{amsmath}
\usepackage{amsfonts}
\usepackage[version=4]{mhchem}

\usepackage[a4paper, total={183mm, 247mm}]{geometry}
\usepackage[switch]{lineno}

\usepackage[superscript,biblabel]{cite}

\newcommand*{\diff}{\mathop{}\!\mathrm{d}}
\newcommand{\ps}{\si{\pico\second}}
\newcommand{\ns}{\si{\nano\second}}
\newcommand{\fs}{\si{\femto\second}}
\newcommand{\ips}{\si{\per\pico\second}}
\newcommand{\us}{\si{\micro\second}} 

\newcommand{\K}{\si{\kelvin}} 
\newcommand{\amu}{\si{\amu}} 
\newcommand{\uzeta}{\si{\pico\second\per\angstrom\squared}}
\newcommand{\THz}{\si{\tera\hertz}}
\newcommand*{\ISF}{\operatorname{ISF}}
\renewcommand*{\Re}{\operatorname{Re}}

\usepackage{sectsty}
\sectionfont{\small}
\subsectionfont{\small}

\title{Justification for the use of Markovian Langevin statistics in the modelling of activated surface diffusion.}
\author{J. Wilkinson$^1$ \and J. Ellis$^1$}

\date{%
	$^1$ Cavendish Laboratory, JJ Thomson Avenue, Cambridge CB3 0HE, UK \\[2ex]%
    \today
}

\begin{document}

\maketitle

\section*{Abstract (195 words)}

Low temperature surface diffusion is driven by the thermally activated hopping of adatoms between adsorption sites. Helium spin-echo techniques, capable of measuring the sub-picosecond motion of individual adatoms, have enabled the bench-marking of many important adsorbate-substrate properties. The well-known Markovian Langevin equation has emerged as the standard tool for the interpretation of such experimental data, replacing adatom-substrate interactions with stochastic white noise and linear friction. However, the consequences of ignoring the colored noise spectrum and non-linearities inherent to surface systems are not known. Through the computational study of three alternative models of adatom motion, we show that the hopping rate and jump distributions of an adatom are fixed to within a few percent by the potential energy surface and a new generalized energy exchange rate parameter alone, independent of the model used. This result justifies the use of the Markovian Langevin equation, regardless of the true statistical nature of adatom forces, provided results are quoted in terms of the new energy exchange rate parameter. Moreover, numerous mechanisms for the effect of noise correlations and non-linear friction on the energy exchange rate are proposed which likely contribute to activated surface diffusion and activated processes more generally. 

\section*{Introduction (571 words)}

Diffusion on the atomic scale is distinguished from the diffusion of larger particles as the potential generated by the medium cannot be considered homogeneous at this scale\cite{Jardine2004}. The regular periodic potential formed by a crystalline substrate results in a form of activated diffusion where adatoms become bound to and hop between local adsorption sites. The motion observed through helium scattering is therefore far more sophisticated than the archetypal diffusion of the Einstein-Smoluchowski type through a homogeneous fluid\cite{Jardine200906}.

The information contained in helium spin-echo measurements is sufficient to benchmark many important properties of the adatom-substrate interaction. These include the activation barrier to diffusion, adatom vibrational frequencies, the rate of hopping between sites, and the strength of atomic scale friction\cite{Jardine200911, Lechner2015, Alexandrowicz, Hedgeland}. These properties are consequential for the development of many technologies and the results of helium spin-echo experiments have contributed to the understanding of graphene growth, self-assembling organic semiconductors, and the design of topological insulators\cite{Tamtgl2015, Townsend, Sacchi, Tamtgl2020}. A thorough understanding of how to model activated diffusion processes and interpret the important quantities involved is therefore of wide interest.
 
The standard approach to modelling diffusive motion uses a Markovian Langevin equation to reduce the forces of a large complex heat bath on a particle of interest to a simple Markovian stochastic force and a dissipative linear friction\cite{Kramers, Zwanzig, Kubo}. This approach is used in the interpretation of almost all helium spin-echo data\cite{Jardine200911, Jardine200906} despite numerous properties of adatom diffusion being in direct conflict with the assumptions used to construct the Markovian Langevin equation.  For instance, the white noise power spectrum implied by the Markovian approximation cannot exist on a real substrate since the dominant source of noise, surface phonons\cite{Rittmeyer2016}, cannot vibrate faster than the Debye phonon cutoff frequency. The cutoff frequency of most lighter metals is around $6-10\THz$ \cite{Sinha, Rao, Zarestky, Stedman1966}, which is not significantly faster than typical adatom vibrational frequencies of $1-4\THz$\cite{Ellis1995, Senet1999LowfrequencyVO, Hofmann1996}. The Debye frequency of heavier metals such as lead and rubidium is found to dip into the low terahertz range\cite{Brockhouse, Copley1973}, making the approximation that adatom motion is effectively driven by white noise manifestly false. The fact that all real forms of noise are band-limited is a negligible detail for most forms of unactivated long-time diffusion\cite{Townsend2018, GlattHoltz2020}. However, even without the complications of atomic scale interactions, some phenomena, such as anomalous sub-diffusion through viscoelastic fluids, can only be accounted for through the use of a generalized non-Markovian Langevin equation\cite{Kubo, GlattHoltz2020, Mason}. The assumption of a linear friction force is also difficult to justify since the exact trajectories of adatoms are not experimentally accessible. From a theoretical perspective, there is no a priori reason for the vanishing of higher powers of velocity in the friction force\cite{Kramers}. The laws of equilibrium thermodynamics can be satisfied by a large number of non-linear friction laws with linear friction being nothing more than the simplest possible case. 

If the forces on an adatom do not resemble Markovian Langevin statistics, the question arises, to what extent are the wide array of results derived from the Langevin equation in activated dynamics meaningful at all? We address this concern through the computational study of alternative models of diffusion which explore the effect of noise correlations and non-linear force laws on activated diffusion. We demonstrate that the details of adatom motion can have a significant effect on hopping behavior, however, only through changing the effective energy exchange rate with the substrate. The results derived through Markovian Langevin simulations are therefore valid provided they are quoted in terms of the new generalized energy exchange rate parameter. 

\section*{Definitions \& computational models (704 words)}

With the notable exception of Hydrogen adsorbates\cite{McIntosh2013}, surface diffusion is predominantly a classical activated diffusion process. The hopping rate and distribution of jump lengths fully specify the rate of macroscopic diffusion over the surface. Both are strongly influenced by the shape of the trapping well, in particular the activation energy, as well as the rate at which the adatom exchanges energy with the substrate. The rate of hopping, $\gamma$, is usually observed to obey an Arrhenius law, of the form $\gamma = \gamma_0 \exp\left(-E_a/k_BT\right)$, with an associated activation energy, $E_a$ and a pre-exponential factor $\gamma_0$. While it is relatively simple to extract an activation energy directly from experimental data\cite{Diamant,Alexandrowicz2006}, it is not possible to disentangle the energy exchange rate of the system from other effects which contribute to the pre-exponential factor without making further assumptions. The most common approach assumes that the force on an adatom of mass $m$ may be treated as a random variable obeying the Markovian Langevin equation in a background potential $U(\vec{r})$,
\begin{equation}
\begin{gathered}
	m\ddot{\vec{r}}=-m\eta\dot{\vec{r}}-\nabla U(\vec{r})+\vec{f}(t) \\ 
	\text{ where } \left<f(t_1)f(t_2)\right>=2k_BTm\eta\delta(t_1-t_2),
	\label{eq:langevin}
\end{gathered}
\end{equation}
and to fit the experimental data with simulated solutions to the Langevin equation. The resulting best fit friction parameter, $\eta$, parameterizes the strength of the random force $\vec{f}$ and may be used as a quantifier of the rate of energy transfer with the substrate. This approach is used in the analysis of almost all helium spin-echo data\cite{Jardine200911, Jardine200906}. 

In addition to the Markovian Langevin equation, we present three models for the motion of adatoms. The first relaxes the Markovian approximation by introducing a correlated noise force with an autocorrelation function given by a memory kernel, $K(t)$. The second fluctuation-dissipation theorem requires that both the friction force and the noise force be modified, resulting in the generalized Langevin equation\cite{Kubo},
\begin{equation}
\begin{gathered}
	m\ddot{\vec{r}}=-m\eta\int_{-\infty}^t\diff{t'}K(t-t')\dot{\vec{r}}(t') - \nabla U(\vec{r}) + \vec{f}(t) \\
	\text{ where } \left<f_i(t_1)f_j(t_2)\right>=2k_BTm\eta K(\left|t_1-t_2\right|)\delta_{ij}.
\end{gathered}
	\label{eq:gle}
\end{equation}
A causal exponential memory kernel, $K(t>0)=\frac{1}{\tau}\exp\left(-\frac{t}{\tau}\right)$ and $K(t<0)=0$, parameterized by a noise correlation time $\tau$ was used to generate a Lorentzian noise power spectrum,
\\
\begin{equation}
	\left<|\tilde{f}_i(\omega)|^2\right> \propto \frac{1}{1 + \omega^2\tau^2}.
\end{equation}
\\
In the limiting case of $\tau=0$, the white noise of the Markovian Langevin equation is recovered. 

The second model explores the effects of non-linear friction through a cubic friction law parameterized by $\zeta$\cite{Kramers},
\begin{equation}
\begin{gathered}
	m\ddot{\vec{r}} = - m\zeta\dot{r}^2\dot{\vec{r}} - \nabla U(\vec{r}) + \vec{f}(t) \text{ where } \\
	\left<f_i(t_1)f_j(t_2)\right>=\left(4\zeta\left(k_BT\right)^2 + 2 m \zeta \left(k_BT\right)\dot{r}^2(t_1)\right)\delta\left(t_1-t_2\right)\delta_{ij}. 
\end{gathered}
\end{equation}
In general, such a friction law may also contain a linear component, however, we primarily present the results of pure cubic friction as an extreme case with no resemblance to the commonly used linear friction. The full linear-cubic friction equation and its associated fluctuation-dissipation relation may be found in work by Kramers\cite{Kramers}.

The final model presented is a full 3D molecular dynamics simulation of sodium on copper(001) which tracks the harmonic interactions and motion of each copper atom in an $8\times8\times8$ substrate as well as the interactions with a sodium adatom via a Morse potential. The simulation parameters determined by Ellis and Toennies were optimized to fit the binding energy, activation energy, and vibrational frequencies of Na on Cu(001)\cite{Ellis}. This model reproduces the phonon dispersion relation of a real copper crystal\cite{Sinha} and therefore provides a realistic model of three dimensional phonon-adatom interactions. Interactions with the surface electron cloud, which have a secondary effect in sodium on copper(001)\cite{Rittmeyer2016}, are not accounted for.

\begin{figure}
	\centering
	\includegraphics[width=1.0\columnwidth]{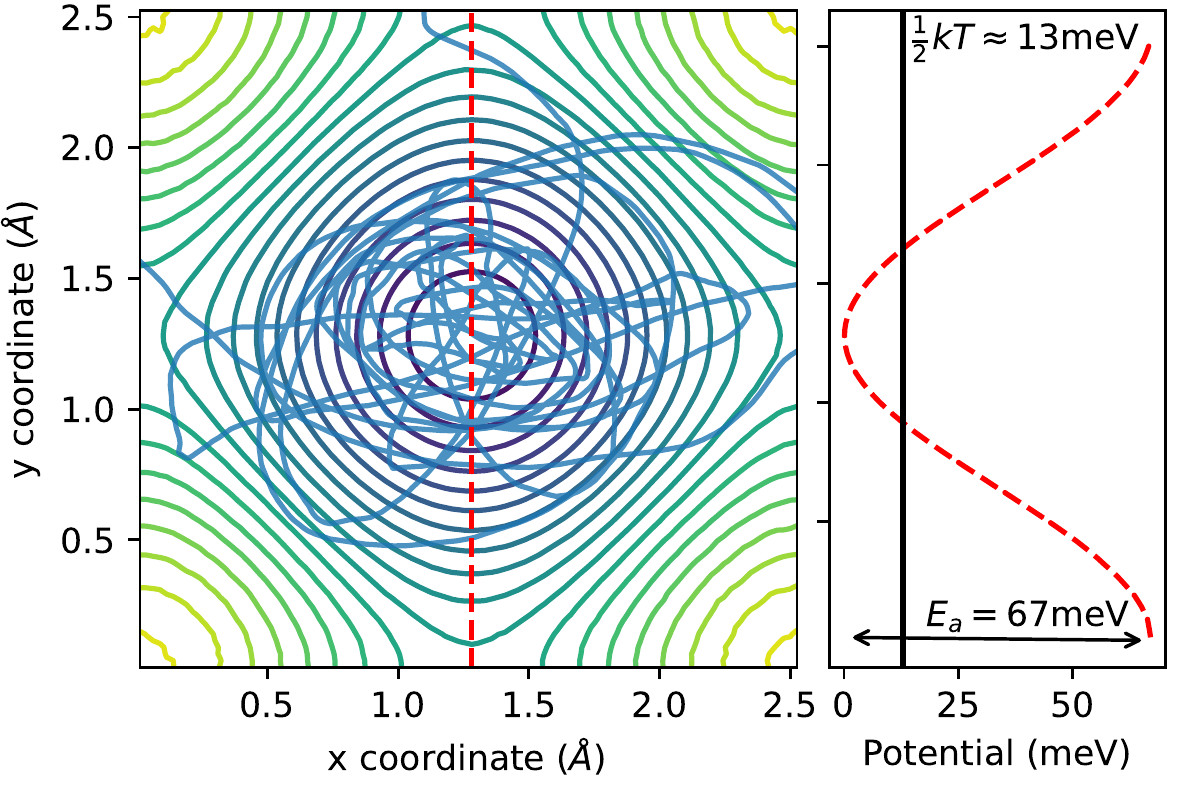}
	\caption{\textbf{Visualization of adatom motion over a potential surface.} The potential energy surface, $U_{\text{free}}$, extracted from the sodium on copper(001) 3D molecular dynamics simulation is shown alongside a superimposed trajectory typical for a low friction activated diffusion process. The red dashed line in the right panel shows a cross-section of the potential through an adsorption site and two bridge sites as annotated in the panel on the left.}
	\label{fig:pot_surface}
\end{figure}

The two dimensional free energy surface parallel to the substrate was extracted from the 3D simulation through the density function of a canonical ensemble of trajectories and the definition (discussed in further detail the methods section),
\begin{equation}
	U_{\text{free}}(\vec{x}) = -k_BT\log\left(\left<\delta^{(2)}(\vec{r}_{\parallel}(t) - \vec{x})\right>\right).
	\label{eq:free_energy}
\end{equation}
The extracted potential, shown in Fig. \ref{fig:pot_surface}, was used as the background potential of the Langevin type simulations and ensures that each model attains the same two dimensional Boltzmannian equilibrium phase space distribution, albeit through fundamentally distinct microscopic statistics. The mass of the adatom in each simulation was set to the mass of sodium, $23\amu$. 

\begin{figure}
	\centering
	\includegraphics[width=1.0\columnwidth]{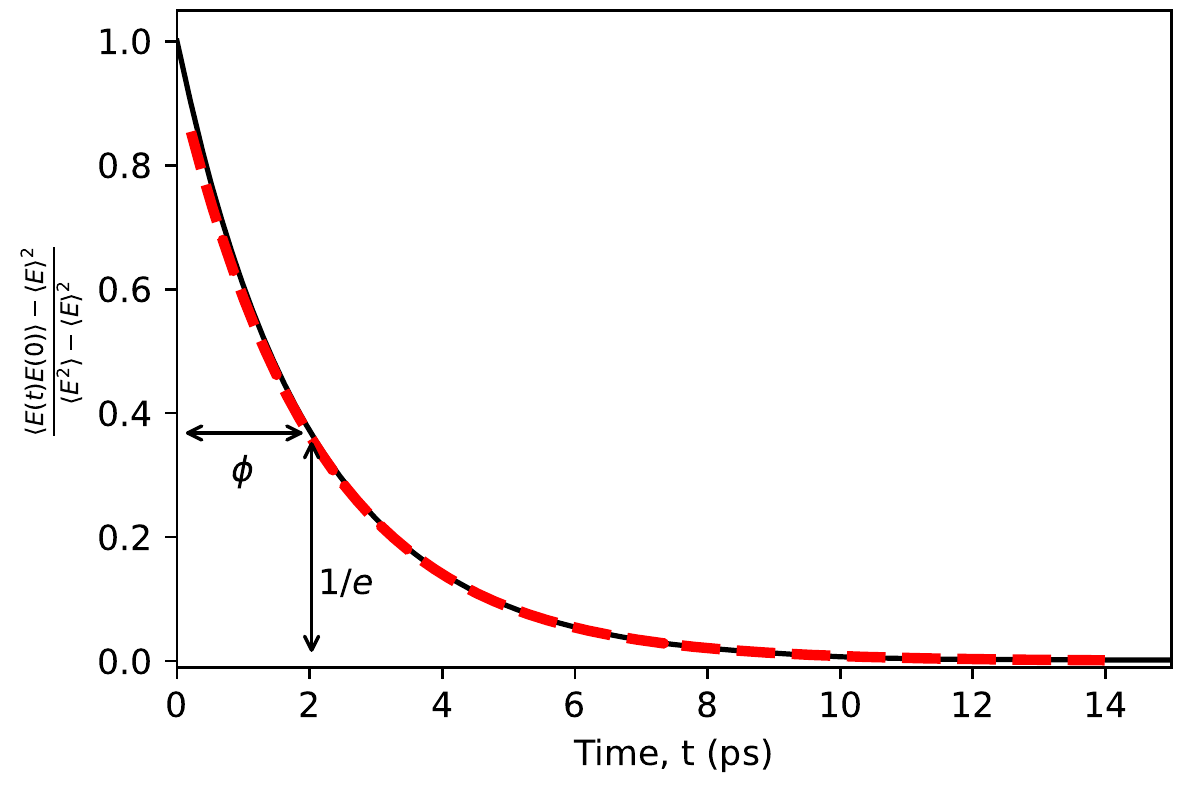}
	\caption{\textbf{Definition of the generalized energy exchange rate, $\phi^{-1}$.} A typical normalized total energy autocorrelation function is shown. Since the total energy autocorrelation function is generally not a pure exponential, as shown by the fit in red, the energy exchange rate, $\phi^{-1}$, is defined as the reciprocal of the time taken for the function to fall by a factor of $\frac{1}{e}$.}
	\label{fig:e_auto}
\end{figure}

The absence of a linear Markovian friction in these alternative models necessitates the introduction of a generalized energy exchange rate parameter. We define the energy exchange rate as the inverse of the correlation time, $\phi$, of the \emph{total energy} autocorrelation function, 
\begin{equation}
	\frac{\left<E(t)E(0)\right> - \left<E\right>^2}{\left<E^2\right> - \left<E\right>^2},
\end{equation}
as defined in Fig. \ref{fig:e_auto}. The energy exchange rate defined through $\phi^{-1}$ has the advantage of being applicable to any model of diffusion while coinciding with $\eta$ in the case of a Markovian Langevin equation in a harmonic well. The total energy autocorrelation function of the 3D simulation contains an additional very rapid decorrelation at short times attributed to interactions with the third coordinate. To compensate for this effect, the decorrelation time of the long-time exponential tail was used in the case of the 3D simulation (further discussion in the supplemental information).

\section*{Hopping rates (789 words)}

\begin{figure}
	\centering
	\includegraphics[width=1.0\columnwidth]{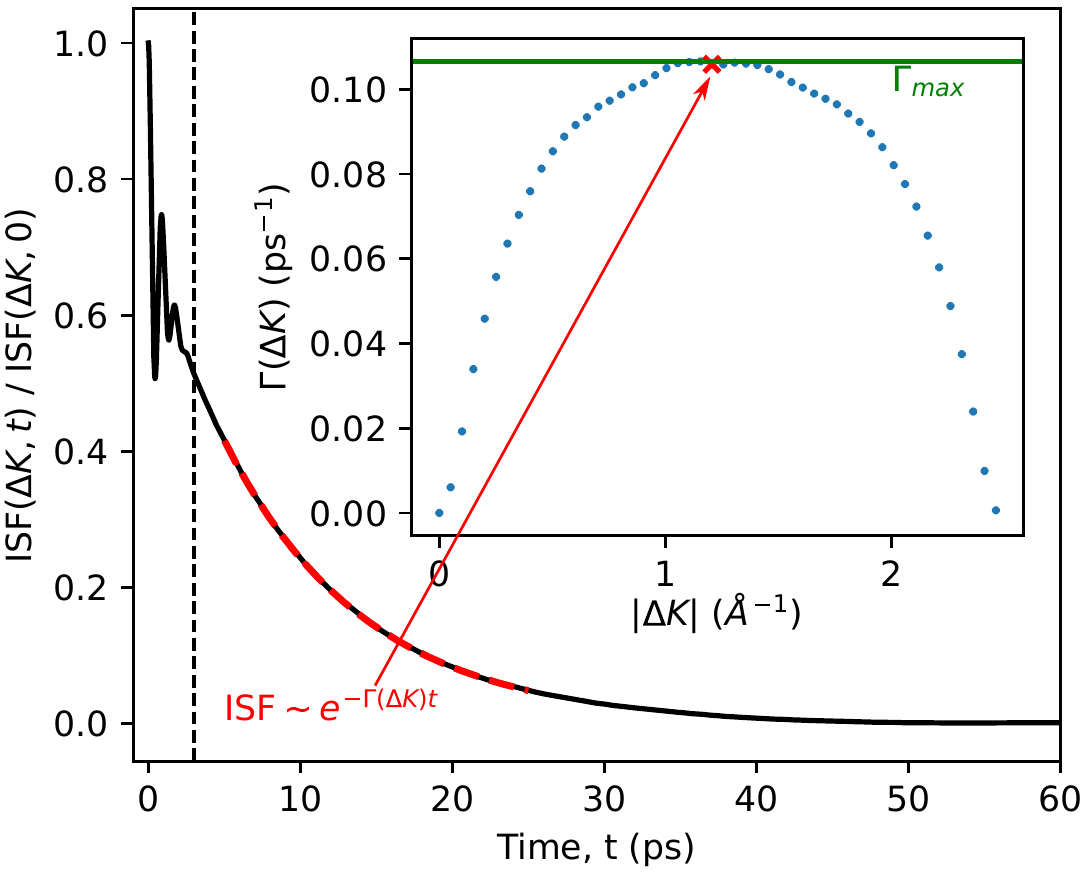}
	\caption{\textbf{Definition of basic surface scattering observables.} The main axes show a typical ISF over time for a fixed momentum transfer. The decay rate, $\Gamma$, of the ISF's exponential tail is proportional to the hopping rate of the adatom. The inset axes show a jump distribution obtained by calculating $\Gamma$ over a range of momentum transfers. The shape of the jump distribution is set by the distribution of adatom jump lengths. All ISFs presented in this paper are quoted with the a momentum transfer pointing from the adsorption site through one of the equivalent bridge sites.} 
	\label{fig:isf_dk}
\end{figure}

Each simulation was run repeatedly at $300\K$ for a total run time of up to $2\us$ per configuration. The hopping behavior of each simulated adatom was evaluated using the observable quantity in a surface scattering experiment, the intermediate scattering function,
\begin{equation}
	\mathrm{ISF}(\Delta{\vec{K}}, t) = \left<\exp\left(i\Delta{\vec{K}}\cdot\vec{r}(t)\right)\right>.
\end{equation}
A typical ISF for a fixed momentum transfer, $\Delta{\vec{K}}$, is shown in the main axes of Fig. \ref{fig:isf_dk}. In the case of activated diffusion, the long time decay rate of the ISF, $\Gamma$, is proportional to the adatom hopping rate, and the shape of $\Gamma$ as a function of $\Delta{K}$, as shown in the inset of Fig. \ref{fig:isf_dk}, is set by the distribution of jump lengths\cite{Chudley, Diamant}.

\begin{figure}
	\centering
	\includegraphics[width=1.0\columnwidth]{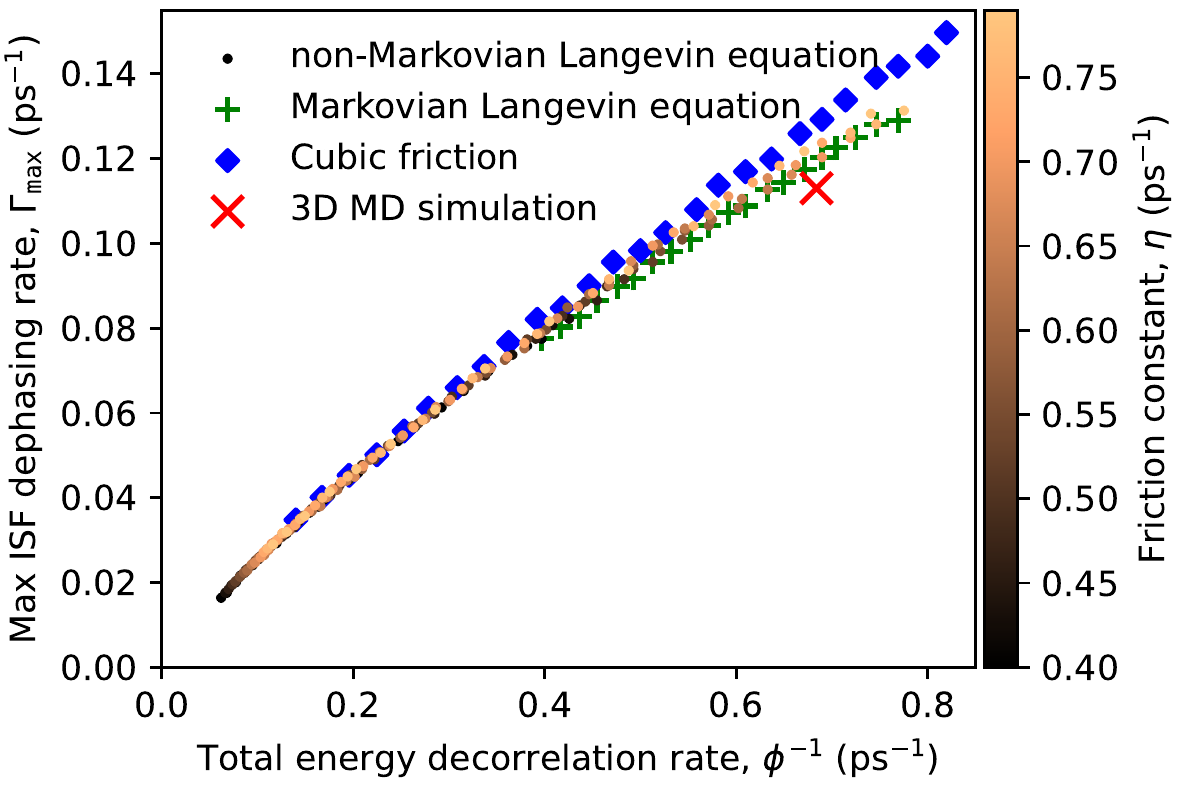}
	\caption{\textbf{A comparison of each model's overall hopping rate dependence on the generalized energy exchange rate, $\phi^{-1}$.} For each model, the maximum ISF dephasing rate, $\Gamma_{\text{max}}$, is plotted against the energy exchange rate, $\phi^{-1}$. The grouping of the models at fixed $\phi^{-1}$ demonstrates that the ISF dephasing rate is predominantly set by the energy exchange rate and the 2D potential background. Other microscopic details account for less than $3\%$ of the variance in the limit of low energy exchange rate. The value of $\eta$ used in the non-Markovian simulation is insufficient to specify the hopping rate, as indicated by the point color.} 
	\label{fig:gamma_ttf}
\end{figure}

\begin{figure}
	\centering
	\includegraphics[width=1.0\columnwidth]{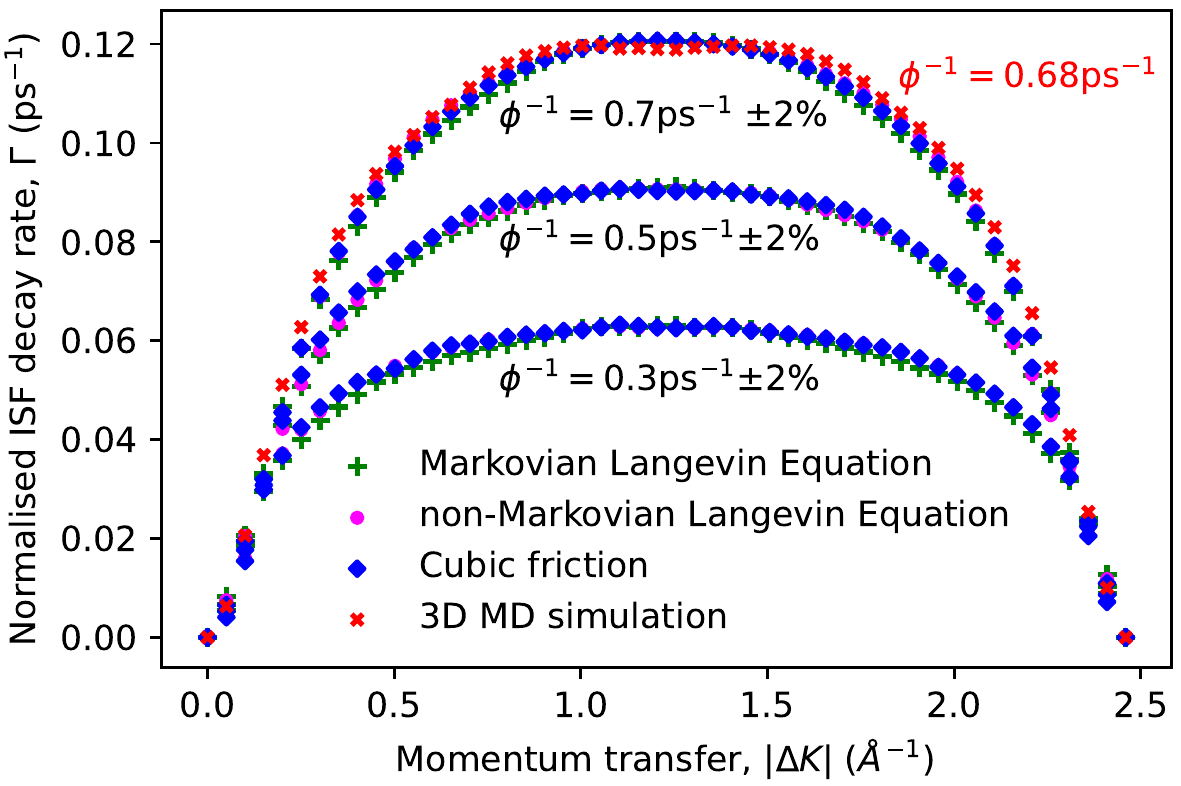}
	\caption{\textbf{A comparison of each model's jump distribution at a fixed generalized energy exchange rate, $\phi^{-1}$.} Each set of points show the jump distribution of a particular model for three energy exchange rates: $\phi^{-1}=0.3$, $0.5$ and $0.7\ips$. A noise correlation time of $\tau=0.1\ps$ was used for the non-Markovian model in each instance. The jump distributions show very little variation across the models at a fixed energy exchange rate.} 
	\label{fig:jump_distribution}
\end{figure}

As a measure of the total hopping rate, the peak of the jump distribution, $\Gamma_{\text{max}}$ (annotated in Fig. \ref{fig:isf_dk}), was calculated for each model and summarized in Fig. \ref{fig:gamma_ttf} as a function of the new energy exchange rate parameter. The results cover almost an order of magnitude of hopping rates as a function of the various model parameters. Despite the wide range of simulated hopping rates, the striking feature of Fig. \ref{fig:gamma_ttf} is that at a fixed energy exchange rate, all adatoms appear to hop at approximately the same rate, regardless of the friction law or noise correlation time used. For instance, the color of non-Markovian model's markers show that, for fixed $\eta$, the total hopping rate can vary by many multiples when the noise correlation time is adjusted. However, when considered as a function of a fixed energy exchange rate, the hopping rate is found to vary by less than $3$\%, regardless of $\tau$. Noise correlations therefore do in fact have an effect on an adatom's hopping rate, but to first order, this effect only occurs through the energy exchange rate and may be compensated for through a corresponding adjustment of $\eta$. Remarkably, there does not appear to be a large behavioral difference between the linear and cubic friction models either. Despite using a fundamentally distinct law of friction, the cubic model's hopping rate was found to follow a very similar trend to that of the linear models as a function of $\phi^{-1}$. The agreement is particularly good in the low energy exchange rate limit where there is no discernible effect of using cubic over linear friction. Even the 3D model, which likely has a unique law of dissipation and contains sophisticated noise correlations was found to hop at a rate within $6$\% of a Markovian Langevin simulation with an analogous energy exchange rate. We conclude that as far as the hopping rate is concerned, the parameter which matters most is the energy exchange rate and the details of how the energy is transferred are mostly inconsequential. 

The shape of the jump distribution of each model was compared by tuning the available parameters to run each model over a fixed set of energy exchange rates. The jump distributions for each model at the energy exchange rate of approximately $0.3$, $0.5$, and $0.7\ips$ are shown in Fig. \ref{fig:jump_distribution}. Since the variation of amplitudes is quantified in Fig. \ref{fig:gamma_ttf}, the jump distributions presented are normalized to peak at the same height within each energy transfer rate to allow for the comparison of the jump distribution shape. For a fixed energy exchange rate, the adatom jump distributions were found to agree well across all models with only small differences in shape. The peaks of the jump distributions at higher exchange rates are all observed to be narrower than those at lower exchange rates, indicating a decrease in the probability of multiple hops\cite{Diamant}. The narrowing effect is expected since higher energy exchange rates decrease the probability that a high energy particle which has escaped its adsorption site finds another transition site before thermalizing. The largest deviation is seen in the 3D model with a greater proportion of adatom jumps terminating outside of the adjacent adsorption sites compared to the 2D models, although the difference is not large.  

These results provide strong evidence that the hopping rate and jump distribution of a low-friction activated diffusion model are fixed by the 2D potential energy surface and the energy exchange rate alone, with almost complete independence of any other details in the low exchange rate limit. The slightly larger deviations of the 3D model are not well understood but are likely due to additional memory effects as a result of energy exchange with the third co-ordinate and the 3D details of the potential affecting the probability of escape when enough energy is obtained. Furthermore the deviations seen across the board are well within the accuracy of current experimental techniques and therefore would not be observable within an experimental context. Evidently, all the results derived from the Markovian Langevin equation so far remain completely valid, not because the underlying system is necessarily described by the Markovian Langevin equation, but rather that the energy exchange rate obtained is correct, regardless of the underlying statistics.

\section*{Energy exchange rates (1069 words)}

\begin{figure*}
	\centering
	\includegraphics[width=0.8\textwidth]{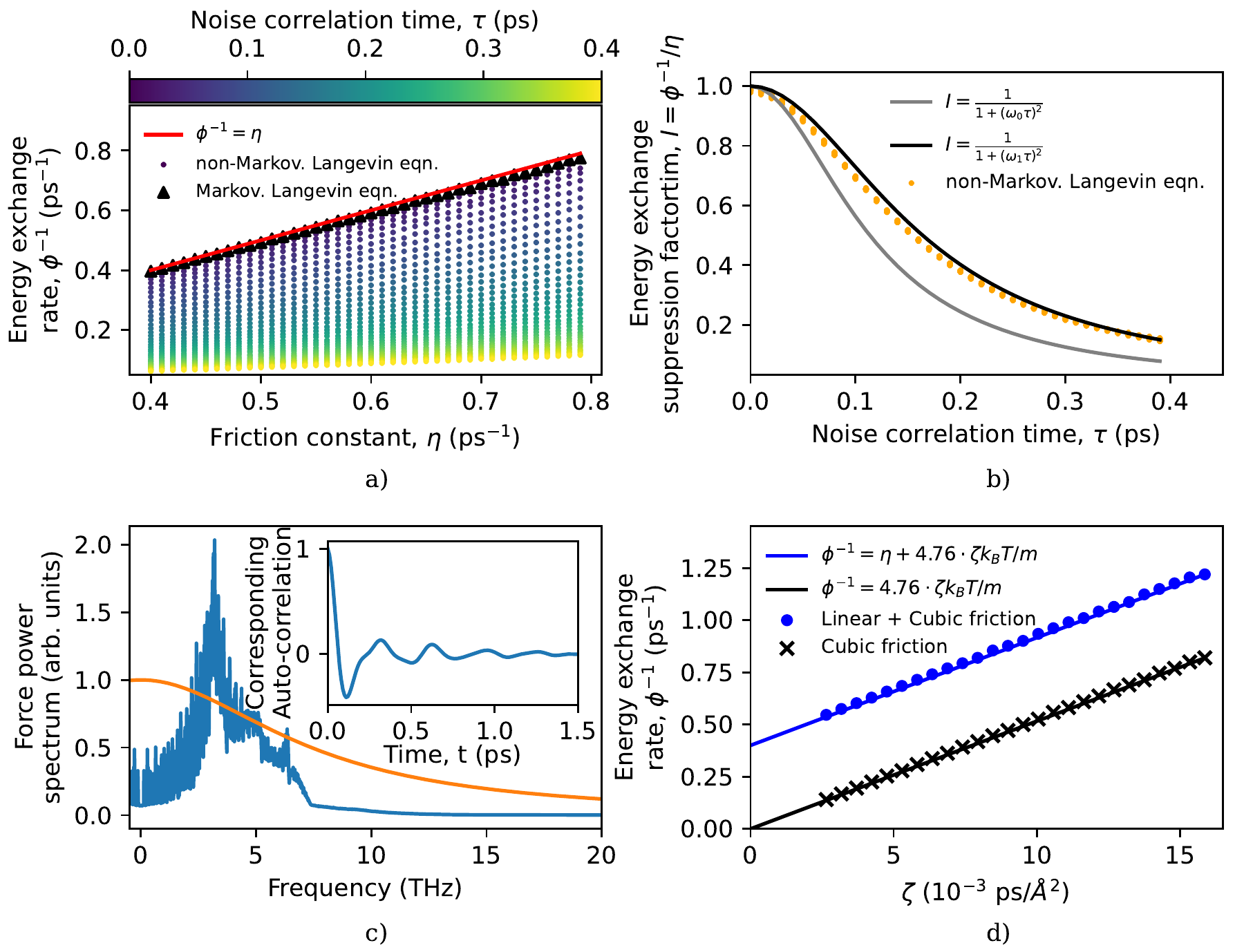}
	\caption{\textbf{Factors influencing the generalized energy exchange rate. a}, the energy exchange rate as a function of the friction parameter and noise correlation time in the linear friction Langevin models. \textbf{b}, the suppression factor, $I=\phi^{-1}/\eta$, of the non-Markovian Langevin model is shown in orange points. The grey line shows shows the suppression factor in a toy model of an equivalent non-Markovian Langevin equation in a harmonic potential of equilibrium natural frequency, $\omega_0=8.8\ips$, matching the curvature of $U_{\text{free}}$. The black line shows the suppression factor of the same toy model with a reduced natural frequency of $\omega_1=6.1\ips$. \textbf{c}, In orange, the form of a Lorentzian power spectrum used in the non-Markovian simulations alongside the power spectrum of the force seen by an adatom held fixed at the minimum of an adsorption site in the 3D simulation. Inset in \textbf{c} is the corresponding time-domain noise autocorrelation function for the force at the equilibrium point of the 3D model. \textbf{d}, in black markers, the energy exchange rate of the cubic friction model as a function of the cubic friction coupling constant. In blue markers, the energy exchange rate of a Langevin simulation run with a combination of cubic and linear friction with $\eta=0.4\ips$. Each solid line shows an expression for the energy exchange rate as as a function of $\eta$ and $\zeta$ obtained through dimensional analysis and a linear fit to determine the proportionality constant $A=4.76$.} 
	\label{fig:energy_exchange_rates}
\end{figure*}

Having established the generalized energy exchange rate as a quantity of interest, we shift focus to evaluating the effects which contribute to $\phi^{-1}$. The introduction of noise correlations was found to suppresses the energy exchange rate with the substrate, as shown in Fig. \ref{fig:energy_exchange_rates}a. For the sake of comparison, the red line shows the energy exchange rate of a particle with Markovian Langevin statistics in a harmonic well, $\phi^{-1}=\eta$. For short correlation times, we find the energy exchange rate is close to $\eta$, but even for $\tau=0$, there are small differences due to anharmonicities in the background potential. As the noise correlation time approaches $0.4\ps$, the energy exchange rate is seen to fall by close to a factor of four, demonstrating that the narrowing of the noise power spectrum can have a large effect on the energy exchange rate. Regardless of the noise correlation time, the exchange rate was found to remain linear in $\eta$ suggesting that $\eta$ continues to act as an overall interaction strength parameter while $\tau$ controls the cutoff frequency of available noise modes. 

The linear dependence of $\phi^{-1}$ on $\eta$ inspires the introduction of a dimensionless, $\eta$ independent, suppression factor $I = \frac{\phi^{-1}}{\eta}$ which quantifies the suppression of energy exchange as a function of $\tau$. The suppression factor for the non-Markovian simulation is shown in Fig. \ref{fig:energy_exchange_rates}b alongside the analytically derived suppression factor of an equivalent non-Markovian particle in a harmonic well evaluated at two natural frequencies, $\omega_0$ and $\omega_1$. The toy model, solved in the supplemental information, is found to admit a similar $\eta$ independent suppression factor in the low friction limit. When the the toy model is evaluated with the natural frequency, $\omega_0$, implied by the curvature of the minimum of $U_{\text{free}}$, there is strong qualitative agreement, however the model appears to underestimate the suppression factor. Interestingly, when evaluated with a natural frequency, $\omega_1$, reduced by $30$\%, there is far better quantitative agreement, likely compensating for some an-harmonicities in background potential. The qualitative agreement of the suppression factors nevertheless indicates a similar underlying origin. The suppression effect can be understood by considering the expected energy change of the adatom when subjected to a single noise impulse which decays in time as $K(t)$. Without loss of generality, we suppose the impulse is imparted in the direction of the instantaneous velocity. If $K(t)$ decays quickly compared to the oscillation period, then the impulse is imparted entirely in the direction of motion, and the adatom quickly settles into a new, higher energy level. If $K(t)$ decays at a rate comparable to the oscillation period of the well then some of the energy imparted initially is removed when the force acts against the direction of motion as the adatom swings back. The new energy level attained is therefore not as high as in the former case and we expect a lower change in the total energy per impulse. A similar argument can be made for the correlated friction force removing less energy per oscillation. This interpretation is reflected in the mathematical form of the low friction approximation of the harmonic well's suppression factor as an oscillating integral over $K(t)$,
\begin{equation}
	I(\tau, \omega_0) = \int_0^{\infty}\diff{t}K(t)\cos{\omega_0t} = \mathrm{Re}\left(\tilde{K}(\omega_0)\right) = \frac{1}{1+\left(\omega_0\tau\right)^2}.
	\label{eq:suppression_factor}
\end{equation}
For an exponential memory kernel, the suppression factor coincides exactly with the power spectrum of the memory kernel, however in general this is not the case. In the supplemental information we show more generally that, in the low-friction limit, the suppression factor in a harmonic well with an arbitrary memory kernel is given by the form of the power spectrum of the noise around $\omega_0$,
\begin{equation}
	I\left(\omega_0\right) = \mathcal{F}\left\{K(\left|t\right|)\right\}(\omega_0) = \frac{\left<|\tilde{f}(\omega_0)|^2\right>}{4\pi k_BTm\eta}.
\end{equation}
Moreover, the noise power spectrum in a physical system is likely to exhibit a much sharper cutoff than the Lorentzian spectrum of an exponential memory kernel. Fig. \ref{fig:energy_exchange_rates}c contrasts the power spectrum of the force experienced by an adatom held stationary at an equilibrium point of the 3D simulation compared to the Lorentzian spectrum used in our non-Markovian simulations. This spectrum is feature rich but clearly demonstrates a sharp cutoff around the copper cutoff frequency of $7.4\THz$. The corresponding time domain noise autocorrelation function contains clear oscillations on the order of a few terahertz. By tuning the mass of the adatom in the 3D simulation, a doubling in the energy exchange rate has been observed when the frequency of the noise correlation oscillations match the adatom's vibrational frequencies, compared to $m=23\amu$. This effect warrants its own investigation but is included here to demonstrate that the effect of noise correlations is likely meaningful in the context of realistic phonon-adatom interactions. High atomic mass substrates with low phonon cutoff frequencies such as lead or rubidium paired with light adatoms may result in the conditions needed to measure systematic changes to the hopping pre-exponential factor when the timescale of adatom vibrational frequencies match features in the the noise spectrum. It is unknown whether the effect would be visible over the other differences in such systems.

Since any realistic non-linear model of friction is likely to contain both linear and non-linear contributions, the energy exchange rate of the cubic model in Fig. \ref{fig:energy_exchange_rates}d is shown alongside a set of simulations run with a combination of linear and cubic friction. The strength of the linear term was set to $\eta=0.4\ips$ which coincides with the y-intercept of the linear fit to the blue points in Fig. \ref{fig:energy_exchange_rates}d. Furthermore, the gradient of $\phi^{-1}$ as a function of $\zeta$ is found to be independent of $\eta$. The implied affine relationship suggests $\phi^{-1}$ may be separated out in a lowest order expansion in the friction constants. Although cubic friction is much more difficult to analyze analytically, simple dimensional analysis on the co-efficient of $\zeta$ in the expansion of $\phi^{-1}$ suggests,
\begin{equation}
	\phi^{-1} = \eta + A \frac{k_BT}{m} \zeta + \text{higher order terms},
	\label{eq:temp_dependence}
\end{equation}
where $A$ is a dimensionless functional of the potential $U_{\text{free}}$. Assuming the form in Eq. \ref{eq:temp_dependence}, Fig. \ref{fig:energy_exchange_rates}d implies $A \approx 4.76$ for the given system. With the exception of background potential anharmonicities, none of the analysis presented up to this point has suggested any temperature dependence of the energy exchange rate. However, in the case of Eq. \ref{eq:temp_dependence}, the temperature is required to account for the dimensions of $\zeta$. Previous work on the Markovian Langevin equation in surface dynamics by Diamant et al. highlighted an anomalous temperature dependence of the linear friction required to fit the same 3D simulation of sodium on copper(001) presented here\cite{Diamant}. We propose that the effect observed by Diamant et al. was in fact a temperature dependent energy exchange rate as a result of cubic contributions to the friction in the 3D model. The data provided by Diamant et al. together with Eq. \ref{eq:temp_dependence} indicate the observed temperature dependence may be achieved using friction constants on the order of $\zeta \approx 6.4\times 10^{-3}\uzeta$ and $\eta \approx 0.37\ips$. With these parameters, the energy exhange rate was found to increase, almost linearly, from $0.46$ to $0.71\ips$ over the temperature range $140-300\K$, comperable to the trend reported by Diamant et al. Moreover, the ratio of the linear to cubic friction terms at thermal velocities is given by
\begin{equation}
	\left<\frac{m\zeta|\dot{\vec{r}}|^3}{m\eta|\dot{\vec{r}}|}\right> = \frac{2k_BT\zeta}{m\eta} \approx 0.23,
\end{equation}
where $\frac{1}{2}m\left<|\dot{\vec{r}}|^2\right>=kT$ in two dimensions, demonstrating a significant portion of the total friction on the sodium adatom may be attributed to cubic friction.  

\section*{Discussion (350 words)}

The results presented re-affirm that the single energy exchange rate parameter obtained through the Markovian Langevin equation is sufficient for describing the transport properties of low friction activated surface diffusion. However, the energy exchange rate must be interpreted as an aggregation of subtle microscopic details, the breakdown of which may not be knowable and may vary as a function of temperature. Although the ability to extract useful information from a system with sophisticated microscopic statistics using a simple Markovian Langevin equation is an interesting result in its own right, the goal of scientific modelling is to construct models which can make causal predictions. This aim warrants a discussion of the features of surface reactions which might be missed in simulations using the Markovian Langevin equation. The most important determinants of the rate of chemical reactions on a surface are the distribution of reagents in position space, how much energy they have, and the rate of replenishment of spent reagents through macroscopic transport over the surface. The results presented demonstrate that a Markovian Langevin equation equipped with the correct energy exchange rate gets precisely this right. For simple monatomic species at low coverage, this is likely the end of the story and the Markovian Langevin equation can likely be safely used for the prediction of surface processes. We however make the final comment that the reactions of larger molecules is influenced by their conformational mobility. Kramers' theory of reaction rates treats the contortion of molecules as a diffusive process in its own right, driven by Markovian noise\cite{Kramers, Zwanzig}. Within Kramers' model, the energy exchange effects we have demonstrated therefore play a role and inspire various possible generalizations of Kramers' low friction escape rate formula. While the timescale of most molecular vibrations is much greater than the phonon cutoff frequency of a substrate, there are examples of certain molecular vibrational modes, in Benzene for example\cite{Wang2020}, which fluctuate as slowly as $10\THz$. If activated states of these vibrational modes are important to the rate of a particular reaction, the details of a substrate's phonon spectrum, not present in a Markovian Langevin simulation, may become relevant. 

\section*{Methods}

\subsection*{Langevin simulations}

The 2D Langevin simulations were performed with a $\Delta{t} = 1\fs$ time step for a total of up to $2\us$ of simulation run time spread over $100-200$ separate $10\ns$ trajectories. At each time step, the background force on the adatom was determined through the gradient of the background potential, as interpolated using a bi-cubic spline on the $100\times100$ grid provided in the Supplemental Information. The random force was sampled independently for each force component from a zero-mean Gaussian random variable with standard deviation $\sigma/\sqrt{\Delta{t}}$ where the relevant fluctuation dissipation relation sets $\sigma$ through $\left<f\left(t_1\right)f\left(t_2\right)\right>=\sigma^2K\left(|t_1-t_2|\right)$. In the non-Markovian simulations, convolution with a discretized exponential memory kernel was performed through the relation
\begin{equation}
	\int_0^{t_n} dt' K\left(t-t'\right) \dot{\vec{r}}(t') \approx \alpha \int_0^{t_{n-1}} dt' K\left(t-t'\right) \dot{\vec{r}}(t') + \frac{1}{1-\alpha} \dot{\vec{r}}\left(t_n\right)
\end{equation}
where $\alpha = \exp{\frac{-\Delta{t}}{\tau}}$. A colored noise spectrum was produced by generating an uncorrelated sequence of impulses and performing an equivalent discretized convolution with $K(t)$, this trick is only valid for an exponential memory kernel. At each time-step, the net force was determined as the sum of the background force, noise force, and the relevant friction law and the adatom position and velocity were propagated forwards in time using a Velocity-Verlet integrator\cite{Verlet}.

\subsection*{3D molecular dynamics simulations}

The 3D molecular dynamics simulation was constructed in a similar fashion to the simulation constructed by Ellis\cite{Ellis} and used the same substrate force constant and Morse potential parameters. The key difference is that a larger $8\times8\times8$ substrate was used and the pairwise adatom interactions with the substrate atoms were tracked up to a separation of $3.1$ times the Morse potential parameter $r_0$ in separation. A Velocity Verlet integrator\cite{Verlet} with a $10\fs$ time step was used to propagate the simulation through time for $200$ runs of $10\ns$ at $300\K$.

\subsection*{Extracting the 2D background potential}

The 3D simulation was used to produce $2\us$ of adatom trajectory at 8 temperatures across the range $140-300\K$. The $2\us$ trajectory of each temperature was mapped back into the first 2D unit cell on the periodic surface and binned into a $100\times100$ 2D grid, ignoring the co-ordinate perpendicular to the substrate. Using Eq. \ref{eq:free_energy}, the 2D free energy was calculated and verified not to vary as a function of temperature. The potential grid provided in the supplemental information and shown in Fig. \ref{fig:pot_surface} is the arithmetic mean of these grids.

\section*{Code availability}

All code was written in a combination of Python and Cython and is available on GitHub through the link \url{https://github.com/jjhw3/gle_research}. In due time I will provide a condensed version of the code as this repository was used for many sub-projects relating to Langevin and Molecular Dynamics simulations and therefore contains some clutter. 

\section*{Data availability}

The trajectories of all 3D molecular dynamics simulations are available upon request. The trajectories of the Langevin simulations were not stored due to storage space constraints, however, equivalent trajectories may be quickly recomputed using $\sim15000$ CPU hours on current computing cluster hardware. The ensemble average of the total energy autocorrelation function and the ISF at $\Delta{K}=1.23\si{\per\angstrom}$ is available for each Langevin simulation upon request. The potential energy surface, $U_{\text{free}}$, is available in the supplemental information.

\section*{Acknowledgements}

We thank the Skye Foundation, Oppenheimer Memorial Trust, and Cambridge Trust for providing the financial support required to perform the work presented in this paper.

This work was performed using resources provided by the Cambridge Service for Data Driven Discovery (CSD3) operated by the University of Cambridge Research Computing Service (www.csd3.cam.ac.uk), provided by Dell EMC and Intel using Tier-2 funding from the Engineering and Physical Sciences Research Council (capital grant EP/P020259/1), and DiRAC funding from the Science and Technology Facilities Council (www.dirac.ac.uk).

\bibliography{bibliography}

\begin{thebibliography}{10}

\bibitem{Jardine2004}
Andrew~P. Jardine, Shechar Dworski, Peter Fouquet, Gil Alexandrowicz, David~J.
  Riley, Gabriel Y.~H. Lee, John Ellis, and William Allison.
\newblock Ultrahigh-resolution spin-echo measurement of surface potential
  energy landscapes.
\newblock {\em Science}, 304(5678):1790--1793, June 2004.

\bibitem{Jardine200906}
A~Jardine, G~Alexandrowicz, H~Hedgeland, W~Allison, and J~Ellis.
\newblock Studying the microscopic nature of diffusion with helium-3 spin-echo.
\newblock {\em Physical chemistry chemical physics : PCCP}, 11:3355--74, 06
  2009.

\bibitem{Jardine200911}
A.P. Jardine, H.~Hedgeland, G.~Alexandrowicz, W.~Allison, and J.~Ellis.
\newblock Helium-3 spin-echo: Principles and application to dynamics at
  surfaces.
\newblock {\em Progress in Surface Science}, 84:323--379, 11 2009.

\bibitem{Lechner2015}
Barbara A.~J. Lechner, Holly Hedgeland, Andrew~P. Jardine, William Allison,
  B.~J. Hinch, and John Ellis.
\newblock Vibrational lifetimes and friction in adsorbate motion determined
  from quasi-elastic scattering.
\newblock {\em Physical Chemistry Chemical Physics}, 17(34):21819--21823, 2015.

\bibitem{Alexandrowicz}
G.~Alexandrowicz, A.~P. Jardine, P.~Fouquet, S.~Dworski, W.~Allison, and
  J.~Ellis.
\newblock Observation of microscopic co dynamics on cu(001) using
  $^{3}\mathrm{H}\mathrm{e}$ spin-echo spectroscopy.
\newblock {\em Phys. Rev. Lett.}, 93:156103, Oct 2004.

\bibitem{Hedgeland}
H.~Hedgeland, P.~Fouquet, A.~P. Jardine, G.~Alexandrowicz, W.~Allison, and
  J.~Ellis.
\newblock Measurement of single-molecule frictional dissipation in a
  prototypical nanoscale system.
\newblock 5(8):561--564, July 2009.

\bibitem{Tamtgl2015}
Anton Tamt\"{o}gl, Emanuel Bahn, Jianding Zhu, Peter Fouquet, John Ellis, and
  William Allison.
\newblock Graphene on ni(111): Electronic corrugation and dynamics from helium
  atom scattering.
\newblock {\em The Journal of Physical Chemistry C}, 119(46):25983--25990,
  November 2015.

\bibitem{Townsend}
Peter Stephen~Morris Townsend.
\newblock {\em Diffusion of light adsorbates on transition metal surfaces}.
\newblock PhD thesis, 2018.

\bibitem{Sacchi}
Marco Sacchi, Pratap Singh, David Chisnall, David Ward, Andrew Jardine, Bill
  Allison, John Ellis, and Holly Hedgeland.
\newblock The dynamics of benzene on cu(111): a combined helium spin echo and
  dispersion-corrected dft study into the diffusion of physisorbed aromatics on
  metal surfaces.
\newblock {\em Faraday Discuss.}, 204, 08 2017.

\bibitem{Tamtgl2020}
Anton Tamt\"{o}gl, Marco Sacchi, Nadav Avidor, Irene Calvo-Almaz{\'{a}}n, Peter
  S.~M. Townsend, Martin Bremholm, Philip Hofmann, John Ellis, and William
  Allison.
\newblock Nanoscopic diffusion of water on a topological insulator.
\newblock {\em Nature Communications}, 11(1), January 2020.

\bibitem{Kramers}
H.A. Kramers.
\newblock Brownian motion in a field of force and the diffusion model of
  chemical reactions.
\newblock {\em Physica}, 7(4):284--304, 1940.

\bibitem{Zwanzig}
R.~Zwanzig.
\newblock {\em Nonequilibrium Statistical Mechanics}.
\newblock Oxford University Press, 2001.

\bibitem{Kubo}
R~Kubo.
\newblock The fluctuation-dissipation theorem.
\newblock {\em Reports on Progress in Physics}, 29(1):255--284, 1966.

\bibitem{Rittmeyer2016}
Simon~P. Rittmeyer, David~J. Ward, Patrick G\"{u}tlein, John Ellis, William
  Allison, and Karsten Reuter.
\newblock Energy dissipation during diffusion at metal surfaces: Disentangling
  the role of phonons versus electron-hole pairs.
\newblock {\em Physical Review Letters}, 117(19), November 2016.

\bibitem{Sinha}
S.~K. Sinha.
\newblock Lattice dynamics of copper.
\newblock {\em Phys. Rev.}, 143:422--433, Mar 1966.

\bibitem{Rao}
R.~Ramji Rao and C.~S. Menon.
\newblock Lattice dynamics, third-order elastic constants, and thermal
  expansion of titanium.
\newblock {\em Phys. Rev. B}, 7:644--650, Jan 1973.

\bibitem{Zarestky}
J.~Zarestky and C.~Stassis.
\newblock Lattice dynamics of \ensuremath{\gamma}-fe.
\newblock {\em Phys. Rev. B}, 35:4500--4502, Mar 1987.

\bibitem{Stedman1966}
R.~Stedman and G.~Nilsson.
\newblock Dispersion relations for phonons in aluminum at 80 and
  300{\textdegree}k.
\newblock {\em Physical Review}, 145(2):492--500, May 1966.

\bibitem{Ellis1995}
J.~Ellis, J.~P. Toennies, and G.~Witte.
\newblock Helium atom scattering study of the frustrated translation mode of co
  adsorbed on the cu(001) surface.
\newblock {\em The Journal of Chemical Physics}, 102(12):5059--5070, 1995.

\bibitem{Senet1999LowfrequencyVO}
Patrick Senet, Jan~Peter Toennies, and Gregor Witte.
\newblock Low-frequency vibrations of alkali atoms on cu(001).
\newblock {\em Chemical Physics Letters}, 299:389--394, 1999.

\bibitem{Hofmann1996}
Frank Hofmann and J.~Peter Toennies.
\newblock High-resolution helium atom time-of-flight spectroscopy of
  low-frequency vibrations of adsorbates.
\newblock {\em Chemical Reviews}, 96(4):1307--1326, 1996.
\newblock PMID: 11848791.

\bibitem{Brockhouse}
B.~N. Brockhouse, T.~Arase, G.~Caglioti, K.~R. Rao, and A.~D.~B. Woods.
\newblock Crystal dynamics of lead. i. dispersion curves at
  100\ifmmode^\circ\else\textdegree\fi{}k.
\newblock {\em Phys. Rev.}, 128:1099--1111, Nov 1962.

\bibitem{Copley1973}
J.~R.~D. Copley and B.~N. Brockhouse.
\newblock Crystal dynamics of rubidium. i. measurements and harmonic analysis.
\newblock {\em Canadian Journal of Physics}, 51(6):657--675, March 1973.

\bibitem{Townsend2018}
Peter S~M Townsend and David~J Ward.
\newblock The intermediate scattering function for quasi-elastic scattering in
  the presence of memory friction.
\newblock {\em Journal of Physics Communications}, 2(7):075011, jul 2018.

\bibitem{GlattHoltz2020}
Nathan~E Glatt-Holtz, David~P Herzog, Scott~A McKinley, and Hung~D Nguyen.
\newblock The generalized langevin equation with power-law memory in a
  nonlinear potential well.
\newblock {\em Nonlinearity}, 33(6):2820--2852, April 2020.

\bibitem{Mason}
T.~G. Mason and D.~A. Weitz.
\newblock Optical measurements of frequency-dependent linear viscoelastic
  moduli of complex fluids.
\newblock {\em Phys. Rev. Lett.}, 74:1250--1253, Feb 1995.

\bibitem{McIntosh2013}
Eliza~M. McIntosh, K.~Thor Wikfeldt, John Ellis, Angelos Michaelides, and
  William Allison.
\newblock Quantum effects in the diffusion of hydrogen on ru(0001).
\newblock {\em The Journal of Physical Chemistry Letters}, 4(9):1565--1569,
  April 2013.

\bibitem{Diamant}
M~Diamant, S~Rahav, R~Ferrando, and G~Alexandrowicz.
\newblock Interpretation of surface diffusion data with langevin simulations: a
  quantitative assessment.
\newblock 27(12):125008, mar 2015.

\bibitem{Alexandrowicz2006}
G.~Alexandrowicz, A.~P. Jardine, H.~Hedgeland, W.~Allison, and J.~Ellis.
\newblock Onset of 3d collective surface diffusion in the presence of lateral
  interactions: $\mathrm{Na}/\mathrm{Cu}(001)$.
\newblock {\em Phys. Rev. Lett.}, 97:156103, Oct 2006.

\bibitem{Ellis}
J.~Ellis and J.P. Toennies.
\newblock A molecular dynamics simulation of the diffusion of sodium on a
  cu(001) surface.
\newblock {\em Surface Science}, 317(1):99--108, 1994.

\bibitem{Chudley}
C~T Chudley and R~J Elliott.
\newblock Neutron scattering from a liquid on a jump diffusion model.
\newblock 77(2):353--361, feb 1961.

\bibitem{Wang2020}
Shaoqing Wang.
\newblock Intrinsic molecular vibration and rigorous vibrational assignment of
  benzene by first-principles molecular dynamics.
\newblock {\em Scientific Reports}, 10(1), October 2020.

\bibitem{Verlet}
Loup Verlet.
\newblock Computer "experiments" on classical fluids. i. thermodynamical
  properties of lennard-jones molecules.
\newblock {\em Phys. Rev.}, 159:98--103, Jul 1967.

\end{thebibliography}


\onecolumn

\section*{Supplemental Information}

\tableofcontents

\section{Energy exchange rate of the 3D simulation}

\begin{figure*}
	\centering
	\includegraphics[width=0.5\textwidth]{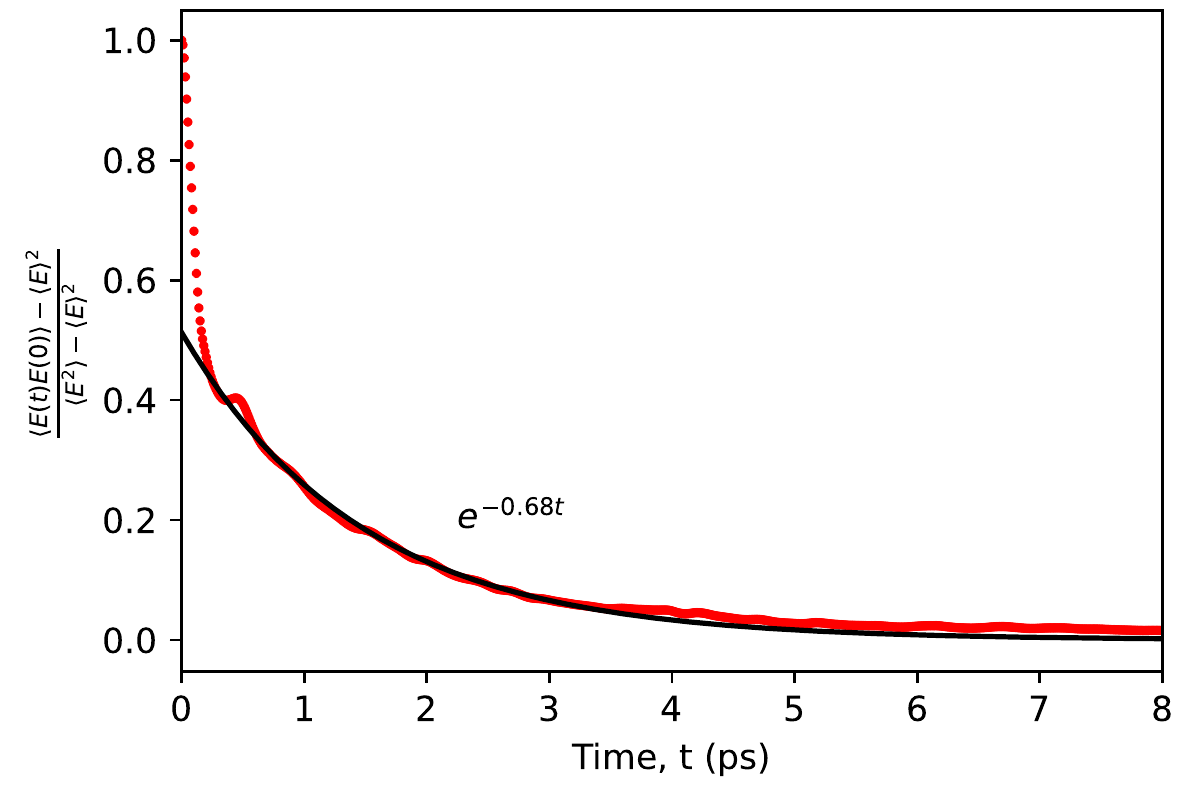}
	\caption{The total energy autocorrelation function of the 3D simulation shows an extremely fast drop at short times not present in 2D simulations. This is assumed to be caused by fast motion in the perpendicular co-ordinate. To compensate for this, the decay rate of the tail is used as the energy exchange rate for this simulation.} 
	\label{fig:MD_e_auto}
\end{figure*}

There is a certain ambiguity as to what energy to use for the definition of the energy exchange rate of the 3D simulation. Although energy may be exchanged with the third co-ordinate, the motion in the this direction is generally much faster than in the lateral directions. This leads to undesirable oscillations in the total energy auto-correlation function if one tries to use a definition in terms of the lateral co-ordinates only. Subsequently, we decided to used the the sum of kinetic energy in all three dimensions and the free energy as a function of all three co-ordinates. The 3D free energy was determined in the same fashion described in the methods section except the position were binned on a 3D grid. The resulting total energy autocorrelation function, Fig. \ref{fig:MD_e_auto}, contains an extremely fast drop at short times not present in 2D simulations. This is assumed to be caused by fast motion in the perpendicular co-ordinate, but is not fully understood. To compensate for this, the decay rate of the tail is used as the energy exchange rate for this simulation. An investigation into how to rigorously define effective 2D properties from such 3D simulations is a potential avenue for future work.

\section{Derivation of the suppression factor for a particle in a harmonic well with non-Markovian noise} \label{sec:suppresion_factor}

The non-Markovian Langevin equation for a particle of mass $m$ in a harmonic well of natural frequency $\omega_0$ is given by
\begin{equation}
	m\ddot{\vec{r}}+m\eta\int\diff{t'}K(t-t')\dot{\vec{r}}(t')+m\omega_0^2\vec{r}=\vec{f}(t)
\end{equation}
\begin{equation}
	\text{ where } \left<f(t_1)f(t_2)\right>=2k_BTm\eta K(\left|t_1-t_2\right|).
\end{equation}
The memory kernel, $K(t)$, is usually assumed to be causal and have a total area of $1$. Taking the Fourier transform of this equation, applying the convolution theorem, and collecting terms results in a solution for the particle's trajectory in Fourier space in terms of a Green's function $\tilde{F}$,
\begin{equation}
	\tilde{x} = \frac{1}{m} \frac{\tilde{f}}{-\omega^2 + i \eta \omega \tilde{K} + \omega_0^2} = \frac{1}{m} \tilde{f} \tilde{F} 
	\label{eq:greens_function}
\end{equation}
\begin{equation}
	\implies x(t) = \frac{1}{m}\int\frac{\diff{\omega}}{2\pi}\tilde{f} \tilde{F} e^{i\omega t}.
\end{equation}
For the sake of clarity, we will make some reasonable claims now which will be justified in following sections. First, we claim that the ensemble average of observables like the energy exchange rate, $\phi^{-1}$, are functions of the positions of the poles of $\tilde{F}$. Furthermore, a non-constant $\tilde{K}$ will in general add additional poles to the Green's function, however, we assume the residue of the pair of poles evident in the denominator of Eq. \ref{eq:greens_function} remain the dominant features of the total energy autocorrelation function. Finally, we will later show that in the low friction limit, the energy exchange rate is given by twice the imaginary co-ordinate of these two poles. We therefore evaluate the energy exchange rate of the system through the shift of these two poles as a result of $\tilde{K}$ by factorizing the denominator in Eq. \ref{eq:greens_function},
\begin{equation}
	\frac{1}{-\omega^2 + i \eta \omega \tilde{K} + \omega_0^2} = \frac{-1}{\left(\omega - \frac{i\eta\tilde{K}}{2} + \omega_0\sqrt{1-\left(\frac{i\eta\tilde{K}}{2\omega_0}\right)^2}\right)\left(\omega - \frac{i\eta\tilde{K}}{2} - \omega_0\sqrt{1-\left(\frac{i\eta\tilde{K}}{2\omega_0}\right)^2}\right)}.
\end{equation}
We proceed by labeling one of the poles $\chi$ and suppose it satisfies
\begin{equation}
	\left(\chi - \frac{i\eta\tilde{K}(\chi)}{2} - \omega_0\sqrt{1-\left(\frac{i\eta\tilde{K}(\chi)}{2\omega_0}\right)^2}\right)=0. \label{eq:def_chi}
\end{equation}
First we note that $\chi=\omega_0$ for $\eta=0$ and proceed by expanding $\chi$ to the lowest order in $\eta$. Implicitly differentiating Eq. \ref{eq:def_chi} with respect to $\eta$ and taking $\eta=0$ demonstrates that $\chi = \omega_0 + \frac{i\eta\tilde{K}(\omega_0)}{2} + O(\eta^2)$. Therefore in the low friction limit, the imaginary component of $\chi$ is shifted to $\frac{\eta\operatorname{Re}(\tilde{K}(\omega_0))}{2}$ and the suppression factor is given by $I=\frac{\phi^{-1}}{\eta}=\frac{2\operatorname{Im}(\chi)}{\eta}=\operatorname{Re}(\tilde{K}(\omega_0))$. In the case of a causal exponential memory kernel with correlation time $\tau$, this gives $I=\frac{1}{1+(\omega_0\tau)^2}$. The co-incidence of $\operatorname{Re}(\tilde{K})$ and $\left|\tilde{K}\right|^2$ is rare and is certainly not the case in general. However, it must be noted that the form of the auto-correlation function of the noise force is given by $K(\left|t\right|)$ where the absolute value signs are critical. Using the Fourier transform of the Heaviside step function (in the sense of a distribution), $\tilde{\theta}(\omega)=\pi\delta(\omega) + i \frac{1}{\omega}$, it is clear that 
\begin{equation}
	\operatorname{Re}\left(\mathcal{F}\left\{K(t)\right\}\right) = \operatorname{Re}\left(\mathcal{F}\left\{K(\left|t\right|)\theta(t)\right\}\right) = \mathcal{F}\left\{K(\left|t\right|)\right\}. 
\end{equation}
In words, the \textbf{Fourier spectrum} (not power spectrum) of $K(\left|t\right|)$ is always equal to $\operatorname{Re}(\tilde{K})$. Since the power spectrum of the noise is given by the Fourier spectrum of $K(\left|t\right|)$, we conclude that the suppression factor in a harmonic well of frequency $\omega_0$ is given by the form of the power spectrum of the noise around $\omega_0$,
\begin{equation}
	I\left(\omega_0\right) = \mathcal{F}\left\{K(\left|t\right|)\right\}(\omega_0) = \frac{\left<|\tilde{f}(\omega_0)|^2\right>}{4 \pi k_BTm\eta}. 
\end{equation}

\section{Observables and Green's function poles}

\subsection{The ISF}

The intermediate scattering function is given by the spatial Fourier transform of the van Hove pair correlation function. For an isolated adatom, this corresponding to the Fourier transform of the position-density function and may be calculated from a single trajectory, $\vec{r}(t)$, using
\\
\begin{equation}
	\ISF(\Delta \vec{K}, t) = \int d\vec{R} e^{-i \Delta \vec{K} \cdot \left(\vec{R} - \vec{r}(0)\right)} P(\vec{R}, t) = \int d\vec{R} e^{-i \Delta \vec{K} \cdot \left(\vec{R} - \vec{r}(0)\right)} \delta(\vec{R} - \vec{r}(t)) = e^{-i \Delta \vec{K} \cdot \left(\vec{r}(t) - \vec{r}(0)\right)}.
	\label{eq:isf_definition}
\end{equation}
\\
Inserting the expression for $\vec{r}(t)$ in Eq. \ref{eq:greens_function} into the definition of the ISF, expanding the exponential, and taking an ensemble average gives
\begin{equation}
	\left<\ISF(\Delta \vec{K}, t)\right> = \sum_{n=0}^{\infty} \left(- \frac{i}{m}\right)^n \frac{1}{n!} \left( \int \frac{d\omega}{2\pi} \left(e^{i\omega t} - 1\right) \tilde{F}\right)^n \left< \left(\Delta \vec{K} \cdot \tilde{f}\right)^n\right>.
	\label{eq:isf_1}
\end{equation}
I have abused notation to summarize $n$ integrals over $\omega_1$ through $\omega_n$. Taking a closer look at $\left< \left(\Delta \vec{K} \cdot \tilde{f}\right)^n\right>$ in one dimension,
\begin{equation}
	\left< \left(\Delta \vec{K} \cdot \tilde{f}\right)^n\right> = \left|\Delta{\vec{K}}\right|^n \left< \tilde{f}\left(\omega_1\right) \ldots \tilde{f}\left(\omega_n\right)\right>.
\end{equation}
From the isotropy and zero mean of $f$ it follows that the above vanishes for odd $n$. Assuming $f$ is a Gaussian random variable, the expectation value for even $n$ is given by the sum over the product of all possible pairwise expectations of $\tilde{f}\left(\omega_1\right) \ldots \tilde{f}\left(\omega_n\right)$. This may be written as a sum over all permutations of pairwise expectations with an appropriate pre-factor dividing out double counting,
\begin{equation}
	\left< \left(\Delta \vec{K} \cdot \tilde{f}\right)^{2n}\right> = \left|\Delta \vec{K}\right|^{2n} \frac{1}{2^nn!} \sum_P \left< \tilde{f}\left(\omega_{P_1}\right) \tilde{f}\left(\omega_{P_2}\right)\right> \ldots \left< \tilde{f}\left(\omega_{P_{2n-1}}\right) \tilde{f}\left(\omega_{P_{2n}}\right)\right>.
\end{equation}

\begin{equation}
= \left|\Delta \vec{K}\right|^{2n} \frac{\left(2\pi\sigma^2\Re(\tilde{K})\right)^n}{2^nn!} \sum_P \delta\left(\omega_{P_1} + \omega_{P_2}\right) \ldots \delta\left(\omega_{P_{2n-1}} + \omega_{P_{2n}}\right).
\end{equation}
The last step follows immediately from the Fourier transform of $\left<f(t)f(t')\right>=\sigma^2 K(|t-t'|)$ and once again abuses notation to summarize the product $\Re(\tilde{K}(\omega_1)) \ldots \Re(\tilde{K}(\omega_n))$. Substituting into \ref{eq:isf_1}, collapsing delta functions, and re-summing the exponential yields,
\begin{equation}
	\left<\ISF(\Delta \vec{K}, t)\right> = \ISF\left(\Delta \vec{K}, \infty\right) \exp\left(\frac{|\Delta \vec{K}|^2 \sigma^2}{m^2} \int \frac{d\omega}{2\pi} e^{i \omega t} \left| \tilde{F}(\omega) \right|^2 \Re(\tilde{K}(\omega)) \right), 
\end{equation}
where $\ISF\left(\Delta \vec{K}, \infty\right)$ is the value of the ISF at long times (in a harmonic well the ISF does not decay to 0), fixed by normalizing $\ISF(\Delta \vec{K}, 0)=1$. Evidently, for $t>0$, the ISF is determined by the reside of the poles of $|\tilde{F}|^2$ in the upper half plane. Since the co-ordinates in a harmonic well are separable and therefore statistically independent, the generalizations to more than one dimension are a trivial consequence of $\left<e^{-i\Delta\vec{K}\cdot\vec{r}(t)}\right>=\left<e^{-i\Delta{K}_1 r_1(t)}\right>\left<e^{-i\Delta{K}_2 r_2(t)}\right>$. 

\subsection{The kinetic energy autocorrelation function}

The velocity the particle may be written in terms of the Green's function as $\dot{x}(t) = \frac{i}{m}\int\frac{d\omega}{2\pi} \omega e^{i\omega t}\tilde{f}(\omega)\tilde{F}(\omega)$. The kinetic energy auto-correlation function is therefore given by
\begin{equation}
	\left<E(0)E(t)\right>=\frac{m^2}{4}\left<\dot{x}(0)^2\dot{x}(t)^2\right>=\frac{1}{4m^2}\left(\int\frac{d\omega}{2\pi}\omega\tilde{F}\right)^4 e^{i\left(\omega_3 + \omega_4 \right)t} \left<\left(\tilde{f}(\omega_1)\cdot\tilde{f}(\omega_2)\right)\left(\tilde{f}(\omega_3)\cdot\tilde{f}(\omega_4)\right)\right>.
\end{equation}
I have once again abused notation to summarize four integrals over $\omega_1\cdots\omega_4$. By expanding the dot product component wise in two dimensions and taking care to sum over all the products of pairwise expectations, $\left<\left(\tilde{f}(\omega_1)\cdot\tilde{f}(\omega_2)\right)\left(\tilde{f}(\omega_3)\cdot\tilde{f}(\omega_4)\right)\right>$ is given by
\begin{equation}
	2\left(2\pi\sigma^2\right)^2\Re(\tilde{K}(\omega_1))\left(2\delta(\omega_1+\omega_2)\delta(\omega_3+\omega_4)\Re(\tilde{K}(\omega_4)) + \delta(\omega_1+\omega_3)\delta(\omega_2+\omega_4)\Re(\tilde{K}(\omega_2)) + \delta(\omega_1+\omega_4)\delta(\omega_2+\omega_3)\Re(\tilde{K}(\omega_3))\right).
\end{equation}
Therefore in 2 dimensions,
\begin{equation}
	\left<E(0)E(t)\right>=\frac{\sigma^4}{m^2}\left(\left(\int\frac{dw}{2\pi}\omega^2\left|\tilde{F}(\omega)\right|^2\Re{\tilde{K}(\omega)}\right)^2 + \left(\int\frac{dw}{2\pi}e^{i\omega t}\omega^2\left|\tilde{F}(\omega)\right|^2\Re{\tilde{K}(\omega)}\right)^2\right).
\end{equation}

\subsection{Explicit formulae for an exponential memory kernel}

\begin{figure*}
	\centering
	\includegraphics[width=0.5\textwidth]{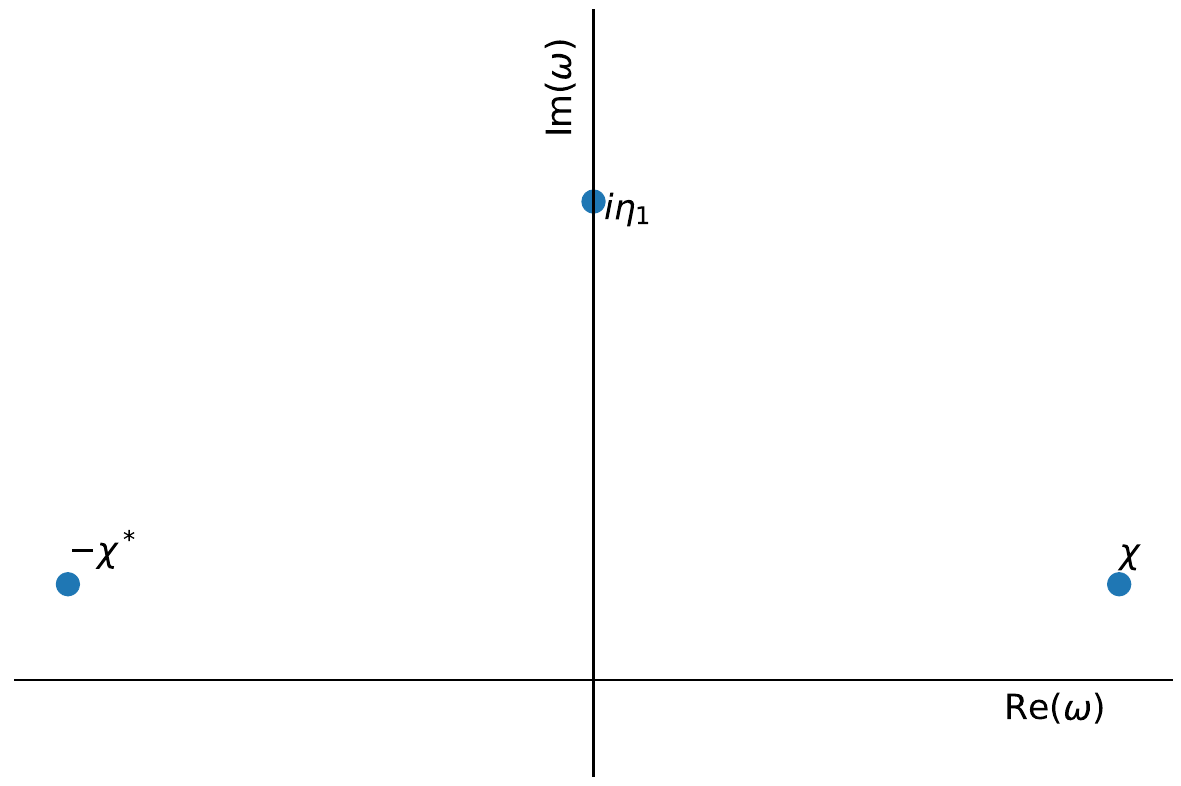}
	\caption{The pole structure of the Green's function of the non-Markovian Langevin equation with an exponential memory kernel.} 
	\label{fig:F_poles}
\end{figure*}

The Fourier spectrum of an exponential memory kernel is given by $\tilde{K}(\omega)=\frac{1}{1+i\omega\tau}$, which results in 3 poles in $\tilde{F}$ given by the zeros of the polynomial
\begin{equation}
	P(\omega) = i\tau\omega^3 + \omega^2 - i\omega(\omega_0^2\tau + \eta) - \omega_0^2.
\end{equation}
Since $P(i\omega)$ is a polynomial with real co-efficients, it follows from the complex conjugate root theorem that for any root $z$ of $P(\omega)$, $-z^*$ is also a root of $P(\omega)$. This allows us to make the ansatz that $P$ may be factorized as $P(\omega) = i\tau\left(\omega - \chi\right)\left(\omega + \chi^*\right)\left(\omega - i\eta_1\right)$ for some $\chi\in\mathbb{C}$ and ${\eta_1\in\mathbb{R}}$. While it is possible that $P(\omega)$ admits three imaginary solutions in some regions of the $\tau,\eta,\omega_0$ parameter space, these parameter ranges have not been encountered in our study of low-friction activated diffusion. If the Greens function $F(t)$ is to be causal, all three roots must occur in the upper half plane. We therefore find that the pole structure of $\tilde{F}(\omega)$ is of the form shown in Figure \ref{fig:F_poles}. A simple application of the residue theorem allows useful integrals of the form $\int\frac{dw}{2\pi} g\left(\omega\right) \left|\tilde{F}\left(\omega\right)\right|^2\Re(\tilde{K}(\omega))$ to be evaluated using the formula,
\begin{equation}
	\int\frac{dw}{2\pi} g\left(\omega\right) \left|\tilde{F}\left(\omega\right)\right|^2 \Re(\tilde{K}(\omega)) =  -\frac{1}{2 \tau^{2}}\left(\frac{1}{2 \chi' \chi''} \operatorname{Re}\left(\frac{g(\chi)}{\chi\left(\chi^{2}+\eta_{1}^{2}\right)}\right)+\frac{g(i\eta_{1})}{\left(|\chi|^{2}+n_{1}^{2}\right)^{2} \eta_{1}}\right), \label{eq:integral_over_CF}
\end{equation}
provided $g(\omega)$ is an entire function which decays sufficiently fast as $\operatorname{Im}({\omega})\rightarrow\infty$. In the formula above, $\chi'$ and $\chi''$ are defined as the real and imaginary components of $\chi$ respectively. Applied to the previously derived expressions of the ISF and kinetic energy autocorrelation function we obtain,

\begin{equation}
	ISF\left(\Delta \vec{K}, t\right) = ISF\left(\Delta \vec{K}, \infty\right) \exp\left(\frac{-\sigma^2\left|\Delta \vec{K}\right|^2}{2m^2\tau^2}\left(\frac{e^{-\chi''t}}{2\chi'\chi''}\operatorname{Re}\left(\frac{e^{i\chi't}}{\chi\left(\chi^2+\eta_1^2\right)}\right) + \frac{e^{-\eta_1t}}{\left(\left|\chi\right|^2+\eta_1^2\right)^2\eta_1}\right)\right) \label{eq:isf_exp}.
\end{equation}

\begin{equation}
	\left<E(0)E(t)\right>=\left<E(0)E(\infty)\right> + \frac{\sigma^4}{4\tau^4m^2}\left(\frac{e^{-\chi''t}}{2\chi'\chi''}\operatorname{Re}\left(\frac{\chi e^{i\chi't}}{\chi^2+\eta_1^2}\right) + \frac{\eta_1e^{-\eta_1 t}}{\left(\left|\chi\right|^2 + \eta_1^2\right)^2} \right)^2 \label{eq:ek_auto}
\end{equation}
\begin{equation}
	\left<E(0)E(\infty)\right> = \frac{\sigma^4}{4\tau^4m^2}\left(\frac{1}{2\chi'\chi''}\operatorname{Re}\left(\frac{\chi}{\chi^2+\eta_1^2}\right) - \frac{\eta_1}{\left(\left|\chi\right|^2 + \eta_1^2\right)^2} \right)^2.
\end{equation}
These expressions are extremely useful as the starting point for certain short-time theoretical calculations and for checking the correctness of computational work. The expression for $\left<E(0)E(\infty)\right>$ is particularly interesting as it is not immediately apparent that it is equal to $(k_BT)^2$. Nevertheless, numerically evaluating $\left<E(0)E(\infty)\right>/(k_BT)^2$ shows extremely small deviations from $1$, less than $10^{-12}$, likely due to numerical error.

\subsection{Energy exchange rate}

Although not a general proof, a careful expansion of the Green's function pole positions in small $\frac{\eta}{\tau}$ demonstrates that the second decay in Eq. \ref{eq:ek_auto} is suppressed by an additional factor of $\left(\frac{\eta}{\omega_0}\right)^2$ compared to the first. Therefore in the low friction limit, the kinetic auto-correlation function is approximately a single oscillating exponential decay with decay rate set by $2\chi''$. Since these two poles are present for all memory kernels (as discussed in Sec. \ref{sec:suppresion_factor}), this first decay is also present for all memory kernels. While more exotic memory kernels may have other decays which dominate the kinetic energy autocorrelation function, these may be numerically evaluated using Eq. \ref{eq:isf_exp} \& \ref{eq:ek_auto}, and we conject the effects are likely to remain small for small $\frac{\eta}{\omega_0}$. Although the total energy autocorrelation function has not been explicitly evaluated, simulated trajectories have shown that the decay rate of the total and kinetic energy autocorrelation functions are the same. We therefore conclude that the non-Markovian Langevin equation has an energy exchange rate in the low friction limit given by $\phi^{-1} = 2\chi''$. 

\end{document}